\documentclass{aa}
\usepackage{graphics,epsfig}

\def\simless{\mathbin{\lower 3pt\hbox
     {$\rlap{\raise 5pt\hbox{$\char'074$}}\mathchar"7218$}}}   
\def\simmore{\mathbin{\lower 3pt\hbox
     {$\rlap{\raise 5pt\hbox{$\char'076$}}\mathchar"7218$}}}   

\begin{document}

\title{Aperiodic variability of low-mass X-ray binaries at very low
frequencies}

\subtitle{}

\author
{P. Reig \inst{1,3}, I. Papadakis\inst{1,2}, N.D. Kylafis\inst{1,2}
}

\institute{
Foundation for Research and Technology-Hellas, GR-711 10 Heraklion, Crete, 
Greece
\and Physics Department, University of Crete, P.O. Box 2208, GR-710 03 
Heraklion, Crete, Greece
\and G.A.C.E, Departament d'Astronomia i Astrof\'{\i}sica, Universitat
de Val\'encia, E - 46071 Paterna-Valencia, Spain
}

\authorrunning{Reig et al.}
\titlerunning{LMXB at very low frequencies}

\offprints{pablo@physics.uoc.gr}

\date{Accepted: \\
Received : \\
}

\abstract{
We have obtained discrete Fourier power spectra of a sample of persistent
low-mass  neutron-star X-ray binaries using long-term light curves from
the {\it All Sky Monitor} on board the {\it Rossi X-ray Timing Explorer}.
Our aim is to investigate their aperiodic variability at frequencies in
the range $1 \times 10^{-7}-5 \times 10^{-6}$ Hz and compare their
properties with those of the black-hole source Cyg X--1. We find that the
classification scheme that divides LMXBs into Z and atoll sources blurs at
very low frequencies. Based on the long-term ($\sim$ years) pattern
of variability and the results of power-law fits
($P(\nu) \propto \nu^{-\alpha}$) to the $1 \times 10^{-7}-5 \times 10^{-6}$ Hz 
power density spectra, low-mass  neutron-star binaries fall into three
categories. Type I includes all Z sources, except Cyg X--2, and the atoll 
sources GX9+1 and GX13+1. They show relatively flat power spectra ($\alpha
\simless 0.9$) and low variability ($rms \simless$ 20\%). Type II systems
comprise 4U 1636--53, 4U 1735--44 and GX3+1. They are more variable 
(20\% $\simless$ $rms \simless$ 30\%) and display steeper power spectra
($0.9 \simless \alpha \simless 1.2$) than Type I sources. Type III systems are 
the most variable ($rms >$ 30\%) and exhibit the steepest power
spectra ($\alpha > 1.2$). The sources 4U 1705--44, 
GX354--0 and 4U 1820--30 belong to this group. GX9+9 and Cyg X--2 appear as
intermediate systems in between Type I and II and Type II and III sources,
respectively. We speculate that the differences in these systems may be
caused by the presence of different types of mass-donor companions. Other 
factors, like the size of the accretion disc and/or the presence of weak 
magnetic fields, are also expected to affect their low-frequency X-ray 
aperiodic varibility.
}

\maketitle

\begin{table*}
\begin{center}
\caption{List of sources}  
\label{soulis}
\begin{tabular}{clllccccccc}
\hline
\hline
Source  &Source	&Alternative	&Class$^a$  &L$_{\rm X}^b$ $\times 10^{38}$&I$_{\rm m}^e$ &$\Delta$I$^f$ &$\alpha^g$ &$\chi^2$/dof$^h$ &$rms$$^i$ &P$_{\rm orb}^j$\\
number  &name	&name		& 		&erg s$^{-1}$		& 	&   		&	&   	&	&hr \\
\hline
\multicolumn{11}{c}{Type I} \\
\hline
1  &4U 1617--15	&Sco X--1	&Z,---,I,2,B	&0.7	&920	&2.1	&0.67$\pm$0.06	&23/24	&14.4$\pm$0.2  	&18.9\\
2  &4U 1642--45	&GX 340+0	&Z,---,II,--,B	&2.2	&30	&2.0	&0.63$\pm$0.05	&43/24	&13.8$\pm$0.2  	&\\
3  &4U 1702--36	&GX 349+2	&Z,Sp,I,2,B	&0.7	&52	&2.0	&0.35$\pm$0.06	&21/24	&15.6$\pm$0.3  	&21.8--22.5\\
4  &4U 1758--25	&GX 5-1 	&Z,Sp,I,--,B	&4.0	&73	&1.6	&0.73$\pm$0.05	&48/23	&12.9$\pm$0.2  	&\\
5  &4U 1758--20	&GX 9+1		&A,Sp,I,2,B	&1.2	&40	&1.7	&0.58$\pm$0.10	&35/23	&10.5$\pm$0.2  	&\\
6  &4U 1811--17 &GX 13+1	&A,Sp,I,--,B	&0.8	&23	&2.0	&0.63$\pm$0.06	&57/23	&11.6$\pm$0.2  	&592.8 \\
7  &4U 1813--14	&GX 17+2	&Z,Sp,I,2,B	&2.3	&47	&1.4	&0.51$\pm$0.06	&31/24	&11.7$\pm$0.2  	&\\
8  &4U 1728--16	&GX 9+9		&A,Su,I,--,B	&0.3	&17	&1.5	&0.85$\pm$0.11	&35/23	&10.3$\pm$0.2  	&4.2\\
\hline
\multicolumn{11}{c}{Type II} \\
\hline
9  &4U 1636--53	&		&A,Su,II,1,D	&0.2	&14	&4.0	&0.97$\pm$0.07	&35/24	&25.9$\pm$0.5  	&3.8\\
10 &4U 1735--44	&		&A,Su,II,--,D	&0.4	&12	&2.4	&0.85$\pm$0.07	&51/24	&22.4$\pm$0.4  	&4.65\\
11 &4U 1744--26	&GX 3+1		&A,Sp,I,--,B	&0.3	&21	&3.0	&1.13$\pm$0.10	&21/23	&22.8$\pm$0.5  	&\\
12 &4U 2142+38	&Cyg X--2	&Z,---,I,--,B	&1.4	&38	&3.7	&1.17$\pm$0.07	&67/25	&24.1$\pm$0.4  	&236\\
\hline
\multicolumn{11}{c}{Type III} \\
\hline
13 &4U 1705--44	&		&A,Su,--,2,D	&0.13	&13	&30	&1.22$\pm$0.12	&76/24	&67.5$\pm$1.2  	&\\
14 &4U 1728--33	&GX 354--0	&A,Su,II,1,D	&0.1$^c$&6	&7.5	&1.79$\pm$0.11	&55/24	&41.7$\pm$0.8  	&\\
15 &4U 1820--30	&		&A,Su,II,--,--	&0.6	&19	&6.0	&1.59$\pm$0.11	&34/24	&32.4$\pm$0.6  	&0.19\\
\hline
\multicolumn{11}{c}{Other sources}\\
\hline
16 &4U 1956+35	&Cyg X--1	&BH		&0.7$^d$&30	&10	&1.04$\pm$0.11	&129/25	&54.0$\pm$0.9 	&134.4\\
\hline
\multicolumn{11}{l}{$a$ Classification of the system in the various
schemes (see text): Z, A=atoll, Sp=supercritical, Su=subcritical} \\
\multicolumn{11}{l}{B=bulge, D=disc, I=class I, II=class II,
1=one spectral component 2=two spectral components BH=black-hole source}\\
\multicolumn{11}{l}{$b$ From Christian \& Swank (1997) in the energy range 
0.5--20 keV} \\
\multicolumn{11}{l}{$c$ From Narita et al. (2001) in the 1--10 keV bandwidth} \\
\multicolumn{11}{l}{$d$ From Belloni et al. (1996) in the 1--30 keV bandwidth} \\
\multicolumn{11}{l}{$e$ Mean ASM count s$^{-1}$ in the energy range 1.3--12.2 keV} \\
\multicolumn{11}{l}{$f$ $\Delta$I=I$_{\rm max}$/I$_{\rm min}$}\\
\multicolumn{11}{l}{$g$ Best-fit power-law index of the power spectra. Errors are the
1$\sigma$ confidence interval} \\
\multicolumn{11}{l}{$h$ $\chi^2$ value and number of degrees of freedom of the 
fits to the power density spectra} \\
\multicolumn{11}{l}{$i$ Fractional (\%) $rms$ amplitude obtained from the light curves as the ratio of 
the variance over the mean} \\
\multicolumn{11}{l}{$j$ From Liu et al. (2001)}\\
\end{tabular}
\end{center}
\end{table*}

\section{Introduction}

A low-mass X-ray binary (LMXB) contains a neutron star which is accreting
material via Roche lobe overflow from a companion star of spectral type
later than A. Due to the high angular momentum  of the accretion flow an
accretion disc is formed around the compact object (see e.g. White 1989).
Unlike high-mass X-ray binaries, whose optical spectrum is dominated by
the emission from the massive companion, in LMXBs the mass-losing star is
usually not seen owing to the contribution of the accretion disc (Cowly
et al. 1991, Shahbaz et al. 1996).  Lewin \& van Paradijs (1985) proposed
that the different classes of LMXBs might reflect the type of companion,
with the brighter systems (usually those located in the galactic bulge)
containing an evolved star and the fainter systems a main-sequence star.
Based on infrared spectroscopic observations, Bandyopadhyay et al. (1999)
suggested that not only the luminosity class but also the spectral type
may be different.

Various schemes have been proposed to categorise the LMXBs:  X-ray
spectral behaviour as a function of intensity  (Parsignault \& Grindlay
1978), cluster analysis of a large number of source characteristics
(Ponman 1982), detailed X-ray spectral fits (White \& Mason 1985), X-ray
hardness-intensity and colour-colour diagrams (Schulz et al. 1988), age
and location in the Galaxy (Naylor \& Podsiadlowski 1993).  In general, a
bimodal distribution of sources and a correlation with luminosity is
found. 

More relevant for the purpose of this paper is the classification scheme
in terms of the rapid aperiodic variability and the patterns that these
sources display in X-ray colour-colour diagrams (Hasinger \& van der Klis
1989). In this scheme LMXBs are divided into two different subclasses,
known as Z and atoll sources.  In the frequency range 10$^{-2}$--10$^3$ Hz
the power spectra of atoll and Z sources are represented by a power law
with index 1--1.5 (atoll sources) and 1.5--2 (Z sources), which describes
the spectrum at low frequencies (below $\sim 1$ Hz) and a power law plus
exponential cut off with index 0--0.8 (atoll sources), $\sim 0$ (Z
sources) and $\nu_{\rm cut}=0.3-25$ Hz (atoll sources), $\nu_{\rm
cut}=30-100$ Hz (Z sources) at higher frequencies (van der Klis 1995 and
references therein). The strength of these two components correlates with
the position of the source in the colour-colour diagram. On top of this
continuum several types of quasi-periodic oscillations are seen (van der
Klis 1994a, 1994b; Wijnands \& van der Klis 1999; Belloni et al. 2002). 
On long time scales ($>$ few tens of days) the studies on LMXB have
concentrated on the search for periodicities (Smale \& Lochner 1992, Kong
et al. 1998). In this work we investigate the long-term aperiodic
variability of 9 atoll and 6 Z persistent LMXBs by comparing the
characteristics of their power density spectra in the frequency range $1
\times 10^{-7}-5 \times 10^{-6}$ Hz and the patterns of variability of
their light curves.

\section{Data analysis}

We have analysed the light curves obtained by the {\em All Sky Monitor}
(ASM) on board the {\it Rossi X-ray Timing Explorer} (RXTE)  of all
persistent neutron-star systems showing an average count rate above 5 c/s
(Table~\ref{soulis}). The data were retrieved from the Definitive Products
Database. The time span by the observations is about 5.7 years, from
February 1996  to October 2001 (JD 2,450,130--2,452,200). The ASM consists
of three wide-angle ($6^{\circ} \times 90^{\circ}$) shadow cameras
(SSC1--3) equipped with position-sensitive Xenon proportional counters
with a total collecting area of 90 cm$^2$. The ASM scans $\sim$ 80\% of
the sky every $\sim$ 90 minutes in a series of dwells of about 90 s each.
Any given X-ray source is observed in about 5--10 dwells every day. The
ASM is sensitive to X-rays in the energy band 1.3-12.2 keV.  For more
information on the ASM see Levine et al. (1996). 

In order to estimate the power density spectrum (PDS) of each source,
we used 1-day binned light curves. Gaps due to detector failure or lack
of data were filled by linear interpolation, adding appropriate random
noise. In order to avoid having a strong dependence on interpolation,
which might introduce undesired features in the power spectrum, data
interpolation was done only when the number of consecutive missing points
was smaller than 2\% of the total number of points in the light curve.  
These short gaps are randomly distributed over the whole light curve. The
average number of missing points per gap was typically 2.5 for the 1-day
rebinned light curves.

The long gaps in the light curves, i.e. those  occurring when no data
points exist during an interval of time longer than 2\% of the total
length of the observation, divide the light curve into segments of
different duration. In order to extract the maximum amount of information
from the light curves we used the discrete (slow) Fourier transform on
each observational segment of data.  Since the data segments had different
lengths, the resulting power density spectra covered different frequency
ranges.  The PDS of the individual segments were merged together, sorted
in order of increasing frequency, and rebinned in frequency to have at
least 40 points per bin. The PDS were white noise  subtracted and
normalized such that their integral over a frequency range $\nu_1-\nu_2$
gives the squared fractional $rms$ variability of the light curves due to
variations on time scales from $\nu_2^{-1}$ to $\nu_1^{-1}$ (Belloni \&
Hasinger 1990, Miyamoto et al. 1991). 

        \begin{figure*}
    \begin{center}
    \leavevmode
\epsfig{file=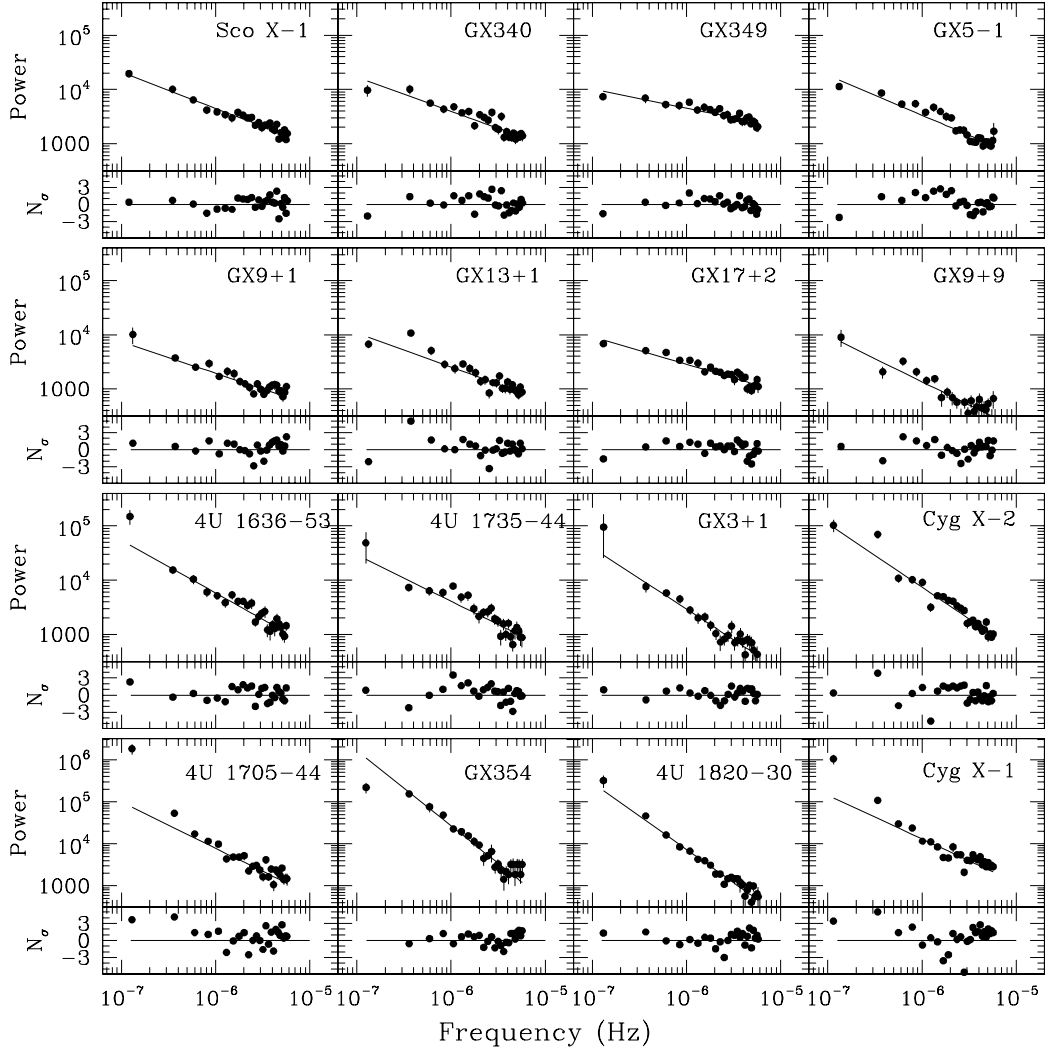, width=14.0cm, bbllx=20pt, bblly=150pt,
  bburx=570pt, bbury=710pt, clip=}
 \end{center}              
        \caption{Power density spectra of the sources studied in this work
and the best-fit power law. The residuals represent the difference between
the observational points and the model, expressed as number of $\sigma$.}
        \label{pds}
        \end{figure*}
        \begin{figure*}
    \begin{center}
    \leavevmode
\epsfig{file=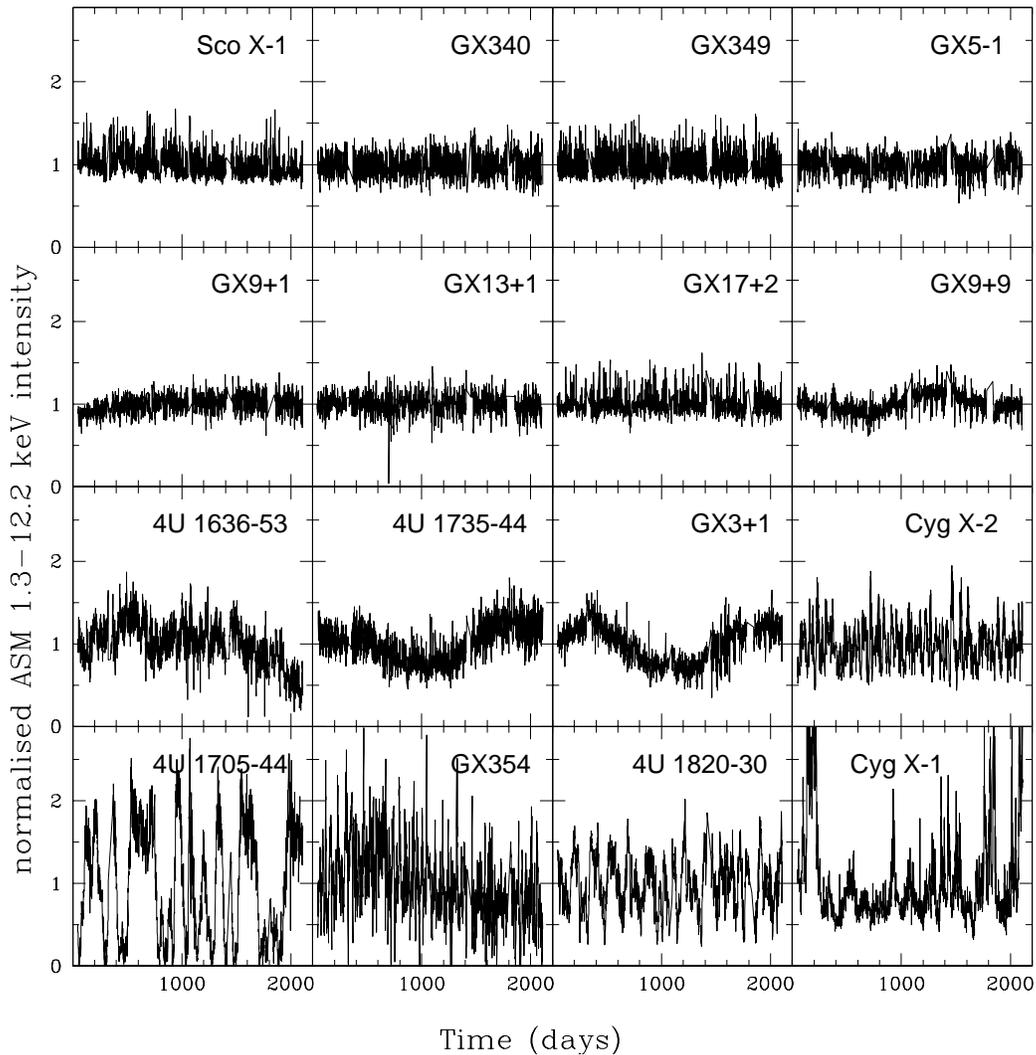, width=14.0cm, bbllx=30pt,
bblly=170pt, bburx=570pt, bbury=710pt, clip=}
 \end{center}              
        \caption{ASM light curves of the sources studied in this work. 
Each point represents 1 day.}
        \label{lc}
        \end{figure*}
        \begin{figure*}
    \begin{center}
    \leavevmode
\epsfig{file=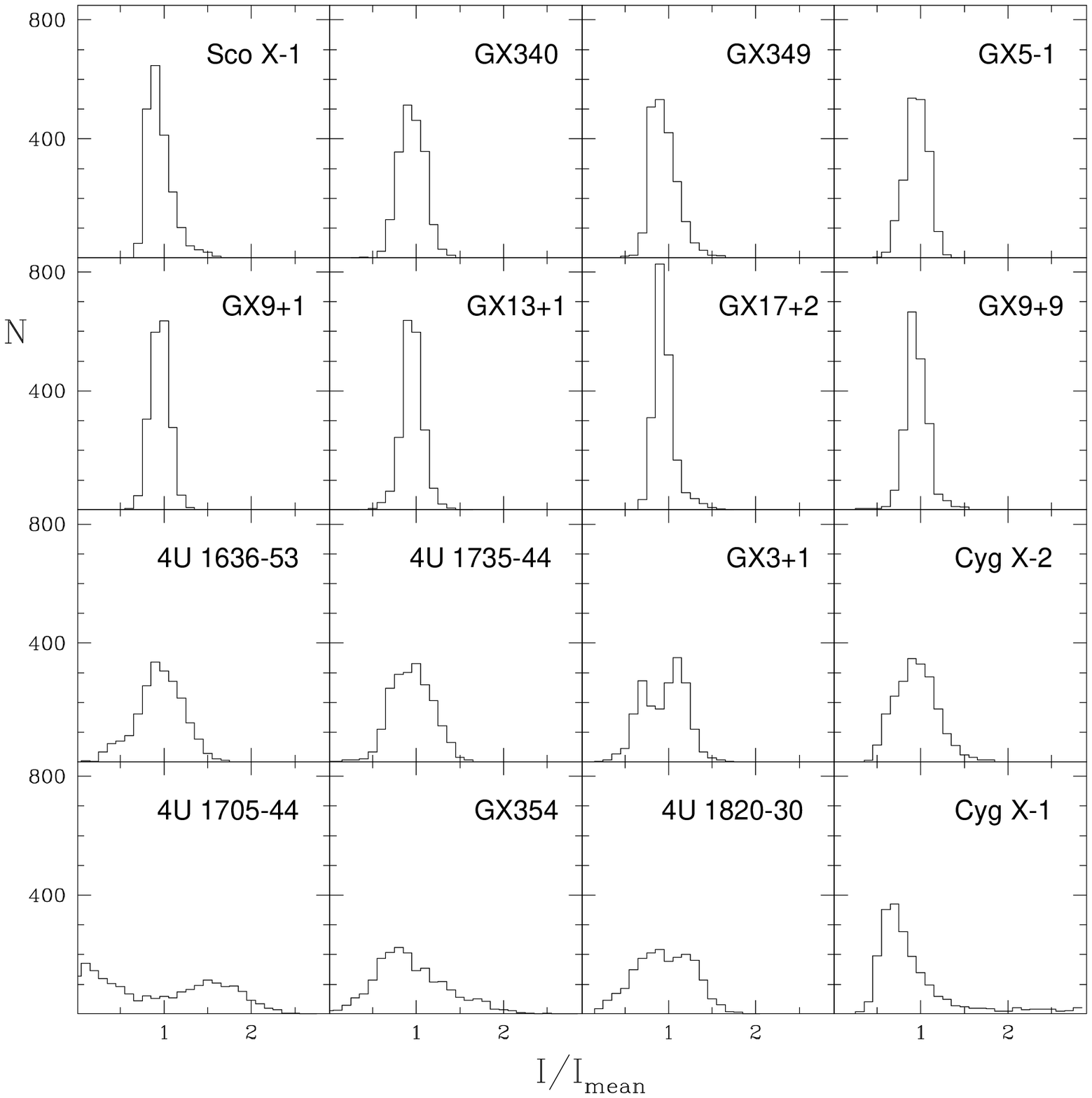, width=14.0cm, 
bbllx=30pt, bblly=170pt, bburx=570pt, bbury=710pt, clip=}
 \end{center}              
        \caption{Distribution of the ASM intensity (normalised to the mean
value) of the sources studied in this work.}
        \label{idf}
        \end{figure*}

\section{Results}

Figure \ref{pds} shows the power density spectra (PDS) of all the
sources studied in this work, together with the best-fit power-law model
and residuals. The residuals represent the number of $\sigma$ that the
observations deviate from the model. For the sake of comparison we have also
analysed the ASM light curve of the black-hole source Cyg X--1 in the same
way (for details on the aperiodic variability of Cyg X--1 at very low
frequencies see Reig et al. 2002).  All sources show a red-noise
dominated power density spectrum (PDS). No flattening at low frequencies
is seen in any of the PDS. Although a single power law did not give
acceptable fits in some cases, it provides a convenient way to compare the
variability of the spectral parameters among the different types of
systems.  Characterising the LMXBs in terms of the shape of the power
density spectrum alone may give rise to ambiguities since intrinsically
different time series can produce similar power density spectra (e.g.
Press 1978). That is to say, two sources may display the same PDS even
though their variability mechanism may {\it not} be the same.  Therefore
we have also examined the X-ray variability of the sources in the time
domain by obtaining the root mean square ($rms$) and the intensity
distribution function (IDF). The $rms$ amplitude gives a measure of the
variability of the source and was computed from the light curve as
$rms=\sigma^2/\bar{x}^2$, where $\bar{x}$ is the mean count rate and
$\sigma^2=\sigma_{\rm obs}^2-\sigma_{\rm exp}^2$ is the difference between
the observed variance, $\sigma_{\rm obs}^2=\sum_{i} (x_i-\bar{x})^2/N$,
and the expected variance, $\sigma_{\rm exp}^2=\sum_{i} \sigma_i^2/N$
($\sigma_i$ are the experimental errors, and $N$ is the total number of
points). Figure \ref{lc} shows the light curves of the sources analysed
here.  The IDFs (Fig.~\ref{idf}) show how often a given (normalised)
intensity occurs. The abscissa represents the intensity of the source
normalised to its mean value in steps of 0.1, whereas the ordinate gives
the number of times (i.e. the frequency) that certain values of the
normalised intensity occur. The IDF plots also give a measure of the
source variability amplitude. Narrow IDFs indicate that the source is
roughly stable, showing low-amplitude variations, whereas highly variable
sources will show many populated bins in their IDF plots, hence resulting
in wide IDFs or narrow IDFs with extended tails.

\begin{figure}
    \begin{center}
    \leavevmode
\epsfig{file=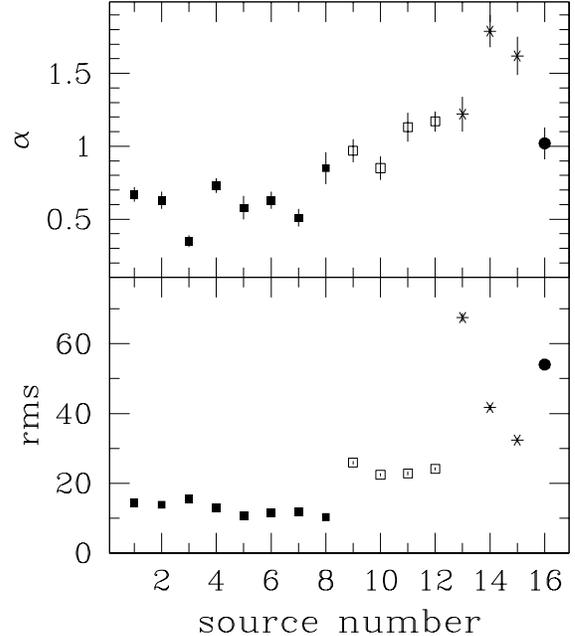, width=8.0cm, bbllx=120pt, bblly=335pt,
  bburx=465pt, bbury=710pt, clip=}
 \end{center}
        \caption{$rms$ amplitude of the light curves and power-law index of 
the power spectra of the individual sources (see Table~\ref{soulis}). 
Type I sources are represented by filled squares, Type II
sources by open squares and Type III sources by stars. The filled circle
corresponds to Cyg X--1. The errors in $rms$ amplitude are smaller than the
symbols.}
        \label{soupar}
        \end{figure}

Figure~\ref{soupar} shows the $rms$ of the light curves and the best-fit PDS
power-law indices. By comparing these power spectral and timing
parameters  LMXBs can be divided into three groups. The first group
includes all Z sources, excluding Cyg X--2, plus the atoll sources GX13+1,
GX9+1 and GX9+9. These are bright sources showing  low-amplitude and fast
variations (on time scales of $\sim$ 1--2 days). As a result, their $rms$
values are small, their PDSs are flat and their IDFs appear to be narrow
and tend to display positive tails. The second group comprises the atoll
sources 4U 1636--53, 4U 1735--44 and GX3+1 and the Z source Cyg X--2.
These sources show similar ASM light curves as the previous group (i.e.
low-amplitude variations on time scales of $\sim 1-2$ days) but with an
extra long-term modulation. This modulation produces more power at low
frequencies giving rise to steeper power spectra and increasing their $rms$
variability. Their IDFs are also broader when compared to the IDFs of the
sources in the previous group and tend to exhibit negative tails. The
third group contains those systems that cannot be placed in neither
of the two previous groups, namely  4U 1705--44, 4U 1820--30 and GX354.
Although they do not form as homogeneous a group as Type I and II sources
they display the highest amplitude variations, the steepest power spectra,
and the broadest IDFs. Quasi-periodic oscillations seem to be a
characterising feature of this group. For the purpose of discussion we
will refer to these groups as Type I, Type II and Type III, respectively.

For comparison, all figures include data of the black-hole source Cyg
X--1. Its PDS slope agrees with that of Type II objects, although it is
also consistent with that of the Type III source 4U 1705-44. Its light
curve and IDF are, however, rather peculiar (note the long tail above
$I/I_{\rm mean}=1.5$). 

\begin{figure}
    \begin{center}
    \leavevmode
\epsfig{file=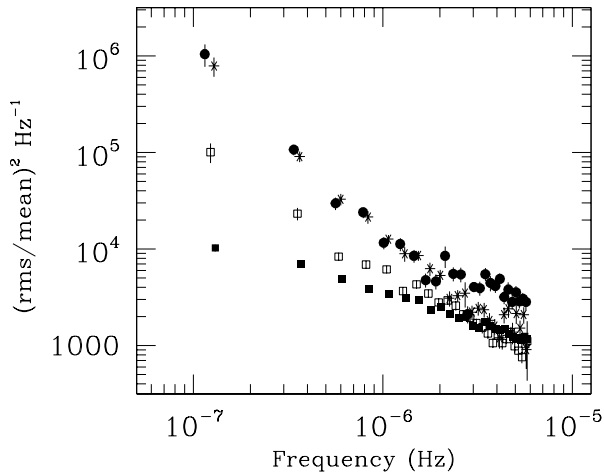, width=8.0cm, bbllx=45pt, bblly=345pt,
  bburx=510pt, bbury=710pt, clip=}
 \end{center}              
        \caption{Comparison of the power spectra of low-mass X-ray binaries
with that of Cyg X--1. The power spectra are
averaged spectra of the individual sources that belong to each group.
Errors are absorbed by the size of the points. Symbols are as in
Fig.~\ref{soupar}.}
        \label{mazi}
        \end{figure}

Since sources belonging to the same group show very similar
characteristics we have obtained a mean power density spectrum for each
group by averaging the PDS of the individual sources pertaining to that
group (Fig.~\ref{mazi}). The PDS of Cyg X--1 has also been plotted (filled
circles). Each group can then be distinguished on the basis of the slope
of the PDS and the amplitude of variability. The power-law index $\alpha$
and the fractional amplitude of variability $rms$ calculated by
integrating over the power-law model in the frequency range $1 \times
10^{-7}-5 \times 10^{-6}$ Hz of the mean PDS for each group of systems
are: $\alpha=0.62\pm0.03$, $rms$=11.5$\pm$0.3\% for Type I sources;
$\alpha=1.09\pm0.05$, $rms$=15.0$\pm$0.4\% for Type II sources and
$\alpha=1.61\pm0.08$, $rms$=29.5$\pm$0.6\% for Type III sources. 

Type I and Type II systems exhibit similar PDS, except for the fact that
Type II sources contain less (more) power at higher (lower) frequencies
than the Type I systems, which agrees with the characteristic shape of the
light curves, i.e. fast variations in Type I LMXB and long-term trends in
Type II LMXB. Type III power spectra show the largest amplitude variations
at all frequencies (except at the highest ones). This
again is in agreement with the flaring like appearance of their light
curves.

Finally, we find a correlation between source luminosity, $L_{\rm X}$, and
$rms$, in the sense that more luminous systems tend to be less variable. In
Fig.\ref{rmslum}, we plot the $rms$ {\em vs} source luminosity. An
anticorrelation between these two quantities is apparent in this figure.
This result is verified when we used the Kendall's $\tau$ nonparametric
statistic (Press et al. 1992) in order to investigate, quantitatively,
whether the apparent anti-correlation between the two quantities is
significant or not. We find $\tau=-0.56$;  the probability that we would
get this value by chance if $L_{\rm X}$ and $rms$ were uncorrelated is $\sim
3.6\times 10^{-3}$. We fitted a straight line to the $\log(L_{\rm X})-
\log(rms)$ plot using the least-squares bisector line method (Isobe
et al. 1990). The results show that $rms\propto L_{\rm X}^{-\beta}$, with
$\beta=0.6\pm0.1$.

\section{Discussion}

We have investigated the aperiodic X-ray variability of all persistent
neutron-star systems showing an average count rate above ASM $RXTE$ 5 c/s
over a period spanning about 5.7 years. A new scheme is proposed in which
LMXBs can be separated into three groups according to the shape of their
power density spectra at very low frequencies and the pattern of the
long-term X-ray variability of their light curves. As classifying
parameters we have taken {\em i)} the fractional amplitude of variability
$rms$ from the light curves, {\em ii)} the slope $\alpha$ of the best-fit
power law from the power spectra and, in a more qualitative way, {\em
iii)} the shape of the intensity distribution functions. 

Type I sources show lower $rms$ and flatter power spectra ($rms < 20\%$, 
$\alpha < 0.9$) than Type II sources ($20\% < rms < 30\%$, $0.9
\simless \alpha \simless 1.2$), and these in turn, have lower $rms$ and
flatter power spectra than Type III ($rms > 30\%$, $\alpha > 1.2$). The
intensity distribution functions of Type I sources tend to show positive
tails and are narrower than those of Type II sources, which tend to show
negative tails. In Type III systems the intensity distribution functions
are even wider.

It should be noted that the use of just one of these parameters does not
generally suffice to characterise unequivocally a system. It is the
combination of the spectral and timing parameters together with the extra
support from the characteristic shape of the intensity distribution
function which allows a more robust way of characterising LMXBs. For
example,  the slope of the PDS in GX9+9 is similar to those found in Type
II sources. However, the $rms$ is about 10 sigmas smaller, more in
accordance with Type I sources as its IDF is. The light curve exhibits a
long-term modulation similar to but lower in amplitude than those of
Type II systems. Likewise, the power-law index of Cyg X--2 is consistent
with the Type III source 4U 1705--44 but its $rms$ amplitude and shape of
the IDF resemble those of Type II sources.

Several schemes have been put forward for the classification of LMXBs:
Parsignault \& Grindlay (1978) separated LMXBs into class I and II
depending on whether or not a direct correlation between the source
temperature (obtained by fitting an exponential to the energy spectrum) and
the source intensity exists. Ponman (1982) distinguished between
subcritical and supercritical LMXBs depending on whether or not the X-ray
luminosity of the system is below or above a certain critical luminosity,
approximately half the Eddington limit. White \& Mason (1985) classified
LMXBs according to the number of components needed to fit their energy
spectra, hence distinguishing between one-component (a single power law
with an exponential cutoff) and two-component (also including a blackbody
function) systems. Based on the age, the location in the Galaxy and
whether or not a low-energy blackbody component was required to fit the
energy spectra Naylor \& Podsiadlowski (1993) divided the LMXBs into bulge
and disc sources. According to the pattern that the source traces out in
the colour-colour X-ray diagram and the type of noise components in the
$10^{-2}-10^{2}$ Hz power density spectra, Hasinger \& van der Klis (1989)
categorised LMXBs into Z and atoll sources. Note that the resulting number
of different types is always two.

In general, the agreement between these classification schemes is good
(see Col. 4 in Table~\ref{soulis}). One group would contain Parsignault
\& Grindlay's class I, Ponman's supercritical, White \& Mason's
two-component spectra, Naylor \& Podsiadlowski's bulge and Hasinger \& van
der Klis' Z sources. The other group would include class II, subcritical,
one-component, disc, atoll sources. Nevertheless, each scheme has some
controversial sources that detach from the norm. For example, GX349+2
(Kuulkers \& van der Klis 1998) and GX13+1 (Homan et al. 1998) exhibit
both Z and atoll characteristics. GX9+9 has a much shorter orbital period
than the rest of its group (bulge sources in the Naylor \& Podsiadlowski's
scheme) making it very unlikely that this system contains a subgiant
secondary as has been proposed for the bulge sources (Bandyopadhyay et al.
1999). GX340+0 did not easily fit in Ponman's scheme (Ponman 1982),
showing characteristics of both, subcritical and supercritical sources. 

When the properties of the long-term variability are taken into account
the classification of LMXBs based on the properties of the aperiodic
variability at higher frequencies is mixed up. For example, Cyg X--2,
which is considered as the prototype of Z sources, shows a much steeper
power spectral continuum ($\alpha=1.2$) and higher amplitude of
variability ($rms$=24\%) than the rest of the Z sources, whose  power-law
index and $rms$ distribute around values 0.7 and 13\%, respectively.
Likewise, atoll sources do not seem to form an homogeneous group either.
Among the atoll sources the power-law index and $rms$ variability vary over
a wide range from 0.6 to 1.6 and 10--60\%, respectively. Some atoll
sources such as GX13+1 or GX9+1 are more similar to Z sources than to
other members of the atoll group. 

It might very well be that the classification of LMXBs in terms of a
discrete number of groups is an oversimplification and that LMXBs
constitute an homogeneous group of sources in which some unknown
parameters change continuously to give rise to their differences.

One of these parameters could be the mass accretion rate. The increase of
the X--ray luminosity from Type III to Type I objects may be interpreted
as an increase in the accretion rate, probably due to the different nature
of the secondary. It was already pointed out by Hasinger \& van
der Klis (1989) that the differences between Z and atoll sources may be
due to the different luminosity class of the mass-donating stars, with Z
sources having evolved secondaries and atoll sources main-sequence
stars.   Infrared spectroscopic measurements (Bandyopadhyay et al. 1999)
show evidence indicating that not only the luminosity class but also the
spectral type might be different so that Z sources would contain evolved
and earlier than G5 stars and atoll sources late-type (K or M), either
evolved or  main-sequence, stars. Since the mass-loss rate scales with
luminosity and radius (Reimers 1975), it is natural to expect that the
mass accretion rate will be higher in systems which have an evolved star
as an optical companion (like the Type I systems).

If the X-ray variations are caused by variations of the accretion rate
which propagate toward the inner regions of the accretion disc, then in
Type I systems, the accretion disc should be able to transfer even the
fastest of these variations (faster even than the diffusion time-scale at
the outer part of the disc). This can be achieved if the accretion disc is
very small in size (perhaps the inner part is missing due to the presence
of a relatively strong magnetosphere), or if it is geometrically thick (e.g.
Churazov et al. 2001).

As the accretion rate decreases, the disc becomes geometrically thin,
allowing only the slowest (and probably largest amplitude) mass-rate
variations to propagate toward the inner part. This could explain the very
slow modulation in the light curves of Type II objects.   Type III objects
could simply be those Type II objects which have: a) the closest companion
(note that the only value of the orbital period known for a Type III
system is much shorter than any other system, see Table~\ref{soulis}) and
b) the weakest magnetic field. In this case, if the accretion disc is
indeed smaller in size (due to the proximity of the companion), then
higher frequency oscillations could also propagate from the outer boundary
toward the inner parts of the disc. At the same time, the larger amplitude
mass accretion rate variations could  penetrate the weak magnetosphere and
extend all the way to the surface of the neutron star, giving rise to the
flaring appearance of their light curves.

\begin{figure}
    \begin{center}
    \leavevmode
\epsfig{file=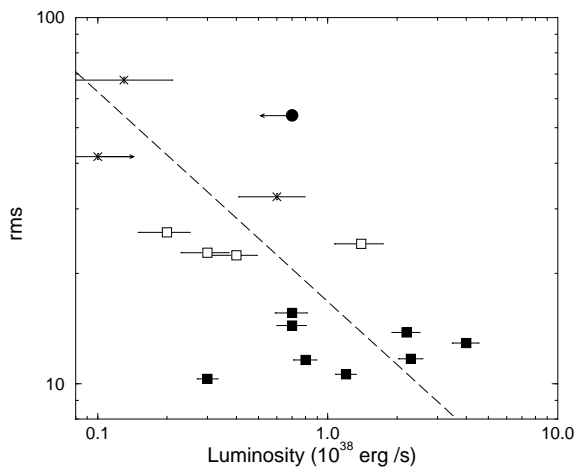, width=8.0cm, bbllx=20pt, bblly=45pt,
  bburx=525pt, bbury=440pt, clip=}
 \end{center}              
        \caption{Fractional amplitude of variability ($rms$) as a 
function of X-ray luminosity in the energy range 0.5-20 keV, except for the
points marked with an arrow, GX354 and Cyg X--1, for which 1-10 keV  and
1--30 keV bandwidths apply, respectively. The dashed line represents the best 
fit to the data using the least-squares bisector line method. Symbols are
as in Fig.~\ref{soupar}.}
        \label{rmslum}
        \end{figure}

This scenario could also explain the dependence of $rms$ on X--ray
luminosity that we observe (Fig.~\ref{rmslum}). Suppose that the observed
variations are indeed caused by variations of the mass accretion rate.
These variations could be thought of as "flares", which originate at the
outermost part of the disc and propagate toward the innermost regions.
Suppose now that the shape and size of the flares are fixed in all
systems, and that only the number of flares per unit time, say $N$, differ
among the various systems. In this case, assuming there is not a
significant constant, underlying component, $L_{\rm X}\propto N$, while
$rms\propto N/L_{\rm X}^{2}$, and so $rms \propto L_{\rm X}^{-1}$, 
inconsistent with the observed relation $rms\propto L_{\rm X}^{-0.6}$.
Therefore, both $N$ {\it and} the typical flare size scale (i.e. duration)
should change with luminosity. If, as we move from Type I to Type II and
Type III objects, only the longest amplitude variations propagate in the
disc (i.e. $N$ decreases) and these variations last longer (i.e. the flare
size scale increases), then a relationship between $rms$ and $L_{\rm X}$,
similar to the one shown in Fig.~\ref{rmslum} could be expected.

It is worth making a comparison between Cyg X--1 and neutron-star systems.
The power spectra of Type I sources are significantly different from that
of Cyg X--1 both in slope and variability. Type II sources, although they
exhibit similar noise shape, are less variable than Cyg X--1. In Type III
sources the $rms$ is comparable to that of Cyg X--1 but have a steeper slope
in the PDS. The differences between the properties of Cyg X--1 and
some of the LMXB such as  4U 1705--44 (whose power-law index and  $rms$
amplitude are similar to those of Cyg X--1) are smaller than between some
LMXBs themselves.  We conclude that the physical processes that cause the
low-frequency X-ray variations in X-ray binaries are determined by
the conditions of matter in the surroundings of the compact object rather
than by its nature (neutron star or black hole).


\section{Conclusions}

Based on the properties of the long-term aperiodic variability we have divided
low-mass X-ray binaries into three categories. Systems in each group
distinguish themselves by the shape and amplitude of variability of  their
power density spectra, light curves and intensity distribution functions.
It seems plausible that Type I systems harbour evolved and possibly early
type companions, whereas Type II and III  systems contain later type
main-sequence stars. The mass loss of the companions increases from Type
III to Type II to Type I systems, hence the corresponding increase in
luminosity. Differences in the accretion disc structure (thin {\em vs}
thick) could explain the lack of fast variations in Type II systems. Type
III systems could also be affected by the weak magnetic field, and perhaps
the small size of the disc. Our results imply that, after all, it is not
just the accretion physics that dictates the observed characteristics of
the systems. Other factors, like the size of the disc and/or the presence
of weak magnetic fields, probably affect the low frequency X-ray
variations in these systems.  Although the separation of LMXBs into these
three groups is useful for the purpose of investigating their general
aperiodic properties, it is likely that LMXBs constitute an homogeneous
group of sources, whose properties vary in a continuous, rather than a
discrete, way. Whether the differences in the aperiodic variability
between Cyg X--1 and LMXBs are due to the different nature of the primary
star -- a B supergiant in opposition to late G-K type stars -- or
intrinsic variations in the vicinity of the compact star is unknown. It
would be interesting to investigate the long-term variability of
black-hole systems with late-type companions. However, owing to the 
transient nature of these systems such studies are very difficult to
perform. 

\begin{acknowledgements} 

PR acknowledges partial support from the European Union via the Training
and Mobility of Researchers Network Grant ERBFMRX/CT98/0195 and from the
Generalitat Valenciana via the Programme Ayudas para las acciones de apoyo
a la investigaci\'on. PR is a researcher of the programme {\it Ram\'on y
Cajal} funded by the University of Valencia and the Spanish Ministery of
Science and Technology. Data provided by the ASM RXTE teams at MIT and at
the RXTE SOF and GOF at NASA's GSFC.

\end{acknowledgements}

\end{document}